\documentclass[usenatbib]{mn2e}
\usepackage{times,graphics,amssymb,epsfig,subfigure,amsmath}
\usepackage{cite,aas_macros,hyperref,url}
\newcommand{\de}{{\rm d}}
\newcommand{\bea}{\begin{eqnarray}}
\newcommand{\eea}{\end{eqnarray}}

\title[Nature of high-$z$ star formation]
{Probing feedback in high-$z$ galaxies using extended UV luminosity functions}
\author[Samui, Srianand \& Subramanian] 
{Saumyadip Samui$^{1}$\thanks{E-mail: saumyadip.physics@presiuniv.ac.in},
        Raghunathan Srianand$^2$\thanks{E-mail: anand@iucaa.in},
	Kandaswamy Subramanian$^2$\thanks{E-mail: kandu@iucaa.in}\\
	$^1$Presidency University, 86/1 College Street, Kolkata 700073, India\\
        $^2$IUCAA, Post Bag 4, Ganeshkhind, Pune 411 007, India.}

\date{Accepted XXX. Received YYY; in original form ZZZ}

\pubyear{2016}

\begin{document}
\label{firstpage}
\pagerange{\pageref{firstpage}--\pageref{lastpage}}
\maketitle

\begin{abstract}
We fit the recently updated UV luminosity functions (LF) of high-$z$ ($1.5\le z\le8.0$)
galaxies using our semi-analytical models of galaxy formation that take into account
various feedback processes. In order to reproduce the overall redshift evolution we
require the efficiency of converting gas into stars to decrease with decreasing
redshift. Even for $z\ge 6$, our models require supernovae (SNe) feedback to reproduce the observed
LF suggesting the prevalence of galactic winds that could have polluted the inter-galactic
medium even at very high redshifts. The observed LF in the low luminosity end for $z<2.5$
shows an upward turn. In our models we reproduce this trend using passively evolving
population of galaxies.  Measuring stellar mass, age and metallicity of these galaxies
using multi-band spectral energy distribution (SED) fitting will place strong constraints on the existence
of such galaxies.  While Active galactic Nuclei (AGN) feedback is  essential to reproduce the LF at high
luminosity end for $z<4$, it may not needed in the case of  $z\ge 6$. We show that the
expected turn around in the LF due to cooling criteria occurs at luminosity much lower than
what has been probed with present day observations. With future deep observations, that can
measure the LF more accurately, we will be able to distinguish between different modes of
SNe feedback and get insights into the physical processes that drive the galaxy evolution.
\end{abstract}

\section{Introduction}

In the most favoured scenario of structure formation today, first structures
hosting galaxies were formed around redshift $z\sim 20 - 30$ \citep{2001PhR...349..125B}.
Stars in these structures produced Ultra-Violet (UV) photons that started
reionizing the intergalactic medium (IGM).
The polarisation of cosmic microwave background radiation (CMBR)
and the Gunn-Peterson effect seen in the $z\ge 6$ quasar spectra can be used to constrain
physical quantities driving this process \citep{2016A&A...596A.108P,2006AJ....131.1203F}. Several
theoretical models of reionisation have been proposed that are consistent with CMBR and quasar
observations \citep{2001PhR...349..125B,2006MNRAS.371L..55C,2007MNRAS.379..253D,
2008ApJ...673L...1G, 2015MNRAS.454L..76M}.
At present consistent reionization can be achieved with UV photons originating
from galaxies (with much higher escape fraction than what is observed in local
galaxies) or using faint quasars \citep{2015ApJ...813L...8M,2016MNRAS.457.4051K}.
A good progress in our understanding of galaxy formation and
various feedback processes at play can be made if we are able to quantify
the faint and bright end slopes of the galaxy luminosity functions at high redshifts.
While the former that gets affected by radiative and supernovae (SNe) feedbacks requires
deep observations, the latter mainly influenced by the Active Galactic Nuclei (AGN)
feedback needs wide field observations.

Indeed, the recent advancement of observational
techniques has enabled us to directly observe the distant faint galaxies
that could have played a major role in the hydrogen
reionisation process. In particular,
with the help of gravitational lensing by the foreground clusters
of galaxies, \citet{2017ApJ...835..113L}
have managed to detect very faint galaxies (upto UV magnitude M$_{\rm UV}=-12.5$)
as early as at $z=8$ when the universe may still be going through the HI reionization.
Taking advantage of the magnification of gravitational lensing they could observe 10
times fainter galaxies compared to direct deep field observations using
Lyman break technique \citep[i.e.][]{2015ApJ...803...34B}.
Their observations show that the faint end slope of UV luminosity function
of galaxies is a power-law that extends upto their observational limits of M$_{\rm UV}=-12.5$
\citep[However see][for discussions on uncertainties in the luminosity function
when lensing magnification is large and when galaxy sizes are unknown]
{2017ApJ...843..129B,2017ApJ...843...41B}.
Further, \citet{2016ApJ...832...56A} presented UV luminosity function
of faint galaxies (i.e. M$_{\rm UV} >-12.5$) at $1<z<3$, using similar
idea of detecting galaxies in the gravitationally lensed fields.
On the other hand \citet{2018PASJ...70S..10O} have measured
the bright end UV luminosity functions of galaxies at
$z \sim 4-7$ upto M$_{\rm UV} = -26$.

The faint end slope of the luminosity function is sensitive to ionisation
feedback \citep[eg.][]{2007MNRAS.377..285S}, presence of massive neutrinos
\citep{2011PhRvD..83l3518J} or warm dark matter \citep{2016ApJ...825L...1M,2017PhRvD..95h3512C}. The above mentioned
observations can play an important role in understanding and/or distinguishing
between these possibilities \citep{2016MNRAS.463.1968Y,2017MNRAS.464.1633F}.
At low and intermediate mass ranges the nature of 
supernovae driven wind feedback can also alter the shape of the
luminosity functions \citep[i.e.][]{2012MNRAS.421.3522H,2014NewA...30...89S,2015ARA&A..53...51S}.
AGN are likely to play
an important role on the amount of star formation in bright galaxies and thereby
influence the bright end of the luminosity function \citep{2006MNRAS.370..645B,2006MNRAS.368L..67B}. 
We have been developing semi-analytical models of galaxy formation to study high-$z$
luminosity functions, galactic outflows and their effects on the IGM
\citep{2007MNRAS.377..285S,2008MNRAS.385..783S,2014NewA...30...89S,2014MNRAS.443.3341J,
2018MNRAS.476.1680S}.
Our models incorporate various feedback processes to reproduce correct shape of
the luminosity function from the dark matter mass function {over a large
  redshift range}. In this work we use the above
mentioned luminosity function observations to place constraints on different
feedback processes in our models and study their redshift evolution.
 
The latest Planck's observations of CMBR suggest a very low optical
depth to the reionisation \citep[$\tau_e=0.058 \pm 0.012$][]{2016A&A...596A.108P}.
They also inferred a rapid reionisation
process with $\Delta z_{re} < 2.8$ at 2$\sigma$ level 
($\Delta z_{re}=z_{10\%}-z_{99\%}$)\footnote{$z_{10\%}$ and $z_{99\%}$
are redshifts when IGM was 10\% and 99\% ionised respectively.}
and reionisation redshift $z_{re}\gtrsim 6$.
Thus we aim to model the observed galaxy luminosity functions at different redshifts
while simultaneously satisfying the 
reionization  constraints from Planck's measurements.

The paper is organised as follows. In the following section we briefly
describe our semi-analytic models. In Sec.~\ref{sec_results} we highlight
our results and finally in section~\ref{sec_dc} we conclude with some
discussions. Through out this work we consider cosmological parameters
that are reported by the Planck's Team \citep{2016A&A...594A..13P},
{\it i.e.} $\Omega_\Lambda = 0.70$, $\Omega_m=0.3$,
$\Omega_b=0.044$, $n_s=0.96$, $H_0=68$~km~s$^{-1}$Mpc$^{-1}$ and $\sigma_8=0.8$ . 

\section{Star formation models}
\label{sec_sf}
We follow our previous star formation model of \citet{2014NewA...30...89S}
that uses Sheth-Tormen halo mass function \citep{1999MNRAS.308..119S},
a simple prescription for star formation and includes supernova
and radiative feedbacks on the star formation in galaxies.
In this model, the star formation rate in a collapsed dark matter
halo evolves with time ($t$) as
\begin{equation}
\frac{d M_*(t)}{dt} = \frac{M_b f_*}{\kappa \tau \eta_w}\Bigg[ 
\exp{\left(-\frac{t}{\kappa\tau}\right)} - \exp{\left(-\frac{(1+\eta_w)t}{\kappa\tau}\right)} \Bigg].
\label{eqn_dM_dt}
\end{equation}
Here, $M_*$ is the mass of stars and $M_b$ is the total baryonic mass that
collapsed with the dark matter and is equal to $(\Omega_b/\Omega_m)M$,
with $M$ being the total halo mass. The star formation efficiency of the halo/galaxy
is governed
by the parameter $f_*$ and the duration of the star formation activity
is govern by the value of $\kappa$; $\tau$ is the dynamical time
scale. Note that $f_*$ is not exactly the star formation efficiency as
traditionally used in the literature. Total mass that would be
finally converted into stars is $M_bf_*/(1+\eta_w)$ \citep[see][for
detailed calculation]{2014NewA...30...89S}.
Further, supernova feedback is regulated by the mass loading factor
$\eta_w$ {which is defined to be the ratio of mass outflowing rate produced
by the supernova explosions to star formation rate}. Depending
on outflow models, $\eta_w \propto v_c^{-2}$ if outflows are energy driven
and/or cosmic ray driven, or $\eta_w \propto v_c^{-1}$ if outflows are momentum driven
\citep[$v_c$ circular velocity of the halo, see][for details]{2008MNRAS.385..783S}.
Further, setting $\eta_w=0$ would produce a star formation scenario
in absence of the supernova feedback (i.e. a close box model of 
star formation). In what follows, we adopt
$\eta_w=(v_c/v_c^*)^{-\alpha}$, $\alpha=2$ and 1 for energy/cosmic ray and momentum
driven outflows respectively, $v_c^*$ is the circular velocity scale at which
$\eta_w=1$.
As the shape of the luminosity function is also regulated by the wind
feedback, one of the motivation of this work is to probe the nature and
extent of the SNe feedback using the observed luminosity functions that spread over
wide luminosity intervals.

Note that the star formation prescription given in Eq.~\ref{eqn_dM_dt}
considers only the onset of a burst of star formation that
resulted due to the formation of the halo. It does not take into account
the passive star formation that originates due to slow accretion
of matter at the later stages of the galaxy evolution as
seen in the red sequence galaxies \citep{2008MNRAS.387...79V}.
In order to model this, we freeze the star formation rate of a galaxy of mass
$M$ after the burst of star formation when it
reaches to a value of $(M/10^{12})$~M$_\odot$/yr. Thus a 
$10^{10}$~M$_\odot$ halo would form stars at a constant rate of 0.01~M$_\odot$/yr
at later stage of its evolution. Such a normalisation
is motivated by the fact that our Galaxy with a halo mass $M\sim 10^{12}$~M$_\odot$
has a constant slow star formation rate of about 1~M$_\odot$/yr \citep{2010ApJ...710L..11R}.
Thus over a 10~billion years of time scale, only $\lesssim 7\%$ of total baryons is converted
to stars due to the adopted model of passive star formation. Note that \citet{2014MNRAS.444.2071D}
obtained such star formation rate in a quasi-steady state ``bathtub" model for
timescales quite longer than the dynamical time \citep[also see][]{2013MNRAS.435..999D}.
In our model the passive mode of star formation starts only after few dynamical times.

The formation rate of dark matter halos at a given redshift, $N(M,z_c)$,
is obtained from the time
derivative of Sheth-Tormen mass function \citep{1999MNRAS.308..119S}
that provides reasonably good fit to the cosmological simulation results.
Note that in principle one could consider the contribution of satellites
in dark matter halos; however, it was shown in \citet{2013MNRAS.429.2333J}
that the contribution of satellite galaxies to the luminosity function is negligibly small
in the redshift range of interest here.
The star formation rate in a given galaxy of mass $M$ that has collapsed
at redshift $z_c$ and being observed at $z$, $z<z_c$, 
is converted to the UV luminosity, $L(M,z,z_c)$, by convolving the luminosity
evolution for a single burst of star formation with a Salpeter initial mass
function in the mass range $1-100$~M$_\odot$
\citep[see][for details]{2007MNRAS.377..285S}.
Further, we assume that the observed luminosity is the intrinsic luminosity reduced by a
factor $\eta$ due to dust reddening. Finally the UV luminosity function,
$\phi(z)$, at a given redshift
is obtained by taking the derivative of cumulative luminosity
function obtained from
\begin{equation}		
\Phi(>L, z ) = \int\limits_z^\infty \de z_c\int\limits_{M_{low}}^\infty
\de M~ N(M,z_c)~ \Theta[L - L(M,z,z_c)].
\label{eqn_lf_cum}
\end{equation}
Here, $\Theta$ is the Heaviside theta function. The lower limit in the mass
integral, $M_{low}$ is decided by the cooling criteria of the gas. In
absence of molecular hydrogen and metals, only gas in halos with virial
temperature above $10^4$~K can cool
and host star formation. Thus we consider $M_{low}$ corresponds to the halo mass
having virial temperature of $10^4$~K \citep{2001PhR...349..125B,2007MNRAS.377..285S}.
Such halos are usually referred to as `atomic cooled halos'.

In our models, in addition to the supernova feedback, we take into account the 
radiative feedback due to meta-galactic UV background after
reionisation and AGN feedback. The former is effective at the low mass
end and the later is effective mainly at the high mass end. We assume that
in the ionised regions of the universe the star formation is completely
suppressed in halos with virial velocity below 35~km/s due to radiative feedback.
For halos having virial velocity between 35~km/s to 110~km/s we assume 
partial suppression in star formation efficiency with a linear fit from 0 to 1
\citep{2002ApJ...575..111B,2002MNRAS.333..156B,2004ApJ...601..666D}.
In the era prior to the epoch of reionisation the radiative feedback is applied
to only galaxies forming in the ionised bubbles. The fraction of such galaxies
at any epoch is quantified by the volume filling factor of the ionised bubbles.
Here we wish to explore whether the currently available observational data are good enough
to constrain the extent to which the radiative feedback is effective. Further,
we consider a suppression of star formation in high mass halos due to
possible AGN feedback by a factor of $[1+(M/10^{12}M_\odot)^3]^{-1}$
\citep{2006MNRAS.370..645B,2006MNRAS.368L..67B}. We also model
the reionisation of inter-galactic medium in a self consistent way
in order to implement the radiative feedback. See \citet{2007MNRAS.377..285S}
for details of such models. All our models predict reionisation histories
consistent with the Planck's measured $\tau_e$.

\section{High redshift luminosity functions}
\label{sec_results}
We first present the UV luminosity functions of high redshift galaxies
as predicted by our models described in previous section
and compare them with the available observations.
This will enable us to constrain
the parameters of our model and hence the nature
of star formation and associated feedback processes at different redshifts.
We vary the star formation efficiency, $f_*$ and
the dust reddening factor $\eta$ together (i.e. $f_*/\eta$ combination) in each redshift
to match our model predictions with observed luminosity functions.
Note that galaxies with the luminosity range $-22<$~M$_{\rm UV}<-20$, are less prone to
any of the feedback processes that we consider here. Hence we choose the value of $f_*/\eta$
such that the model prediction matches with observation in this luminosity range.

The best fit values of $f_*/\eta$ at different redshifts are given
in Table~\ref{tab_param}
for the model with $\eta_w=(v_c/100~{\rm km/s})^{-2}$ and $\kappa=4$.
It shows a decreasing trend of $f_*/\eta$ with decreasing redshifts
\citep[][also got similar results albeit using the models without SNe
feedback]{2014MNRAS.443.3341J}.  This
could arise either from the decreasing star formation efficiency
with time  (i.e $\eta$ remains constant) or time evolution in both $f_*$ and $\eta$.
Note that it has been found that
dust attenuation can decrease (i.e. decreasing $\eta$) with increasing redshifts
\citep{2012ApJ...754...83B,2013A&A...554A..70B,2015ApJ...805...33K}.
In order to explore it in more details, we use the fitting formula as obtained
by \citet[][Eqn.~15]{2018arXiv180109693K} to calculate the dust opacity ($\eta$) at different
redshifts. Using that we calculate the $f_*$ from the fitted values of $f_*/\eta$. However,
it should be noted that this $f_*$ is not the total star formation efficiency as used in literature.
Rather, it determines $M_*/M_b=f_*/(1+\eta_w)$. In Table~\ref{tab_param} we provide
the value of $M_*/M_b$ at halo masses (also mentioned in 2nd row of the table) for which
$\eta_w=1$. This indicates that the total mass converted into stars also
decreases with decreasing redshift.
Below we provide a detailed comparison of our predicted luminosity functions with observations
at different redshifts.

\begin{table*}
\begin{center}
\begin{tabular}{|c|c|c|c|c|c|c|c|c|c|} \hline \hline
$z$ & 8 & 7 & 6 & 5 & 4 & 3 & 2.5 & 1.9 & 1.5 \\ \hline
$f_*/\eta$         & 0.756 & 0.574 & 0.397 & 0.330 & 0.250 & 0.144 & 0.140 & 0.075 & 0.057 \\ \hline
$\eta^\dagger$     & 1.54  & 1.67  & 1.86  & 2.16  & 2.68  & 3.67  & 4.48  & 5.76  & 6.64  \\ \hline
$M^{\ddagger}$     & 2.4   & 2.9   & 3.6   & 4.5   & 5.9   & 8.2   & 10.0  & 13.1  & 16.2 \\ \hline
$M_*/M_b^\mathsection$  & 0.58  & 0.48  & 0.37  & 0.35  & 0.33  & 0.26  & 0.31  & 0.22  & 0.19 \\ \hline \hline
\multicolumn{10}{l}{${}^\dagger$ Calculated using Eqn.~15 of \citet{2018arXiv180109693K}.}\\
\multicolumn{10}{l}{${}^\ddagger$ Mass in unit of $10^{10}$~M$_\odot$ for which $v_c=100$~km/s.}\\
\multicolumn{10}{l}{${}^\mathsection$ Ratio of stellar to baryon mass for halos with $\eta_w=1$.}
\end{tabular}
\caption{The fitted values of $f_*/\eta$ at different redshifts for our fiducial
model with $\alpha=2$ and $\kappa = 4$. We also provide the ratio of $M_*/M_b$ at halo masses with
$\eta_w=1$ by using the value of dust reddening correction ($\eta$) obtained from \citet{2018arXiv180109693K}.}
\label{tab_param}
\end{center}
\end{table*}

\subsection{UV Luminosity functions at $z\ge 6$}
\begin{figure*}
\centerline{
\epsfig{figure=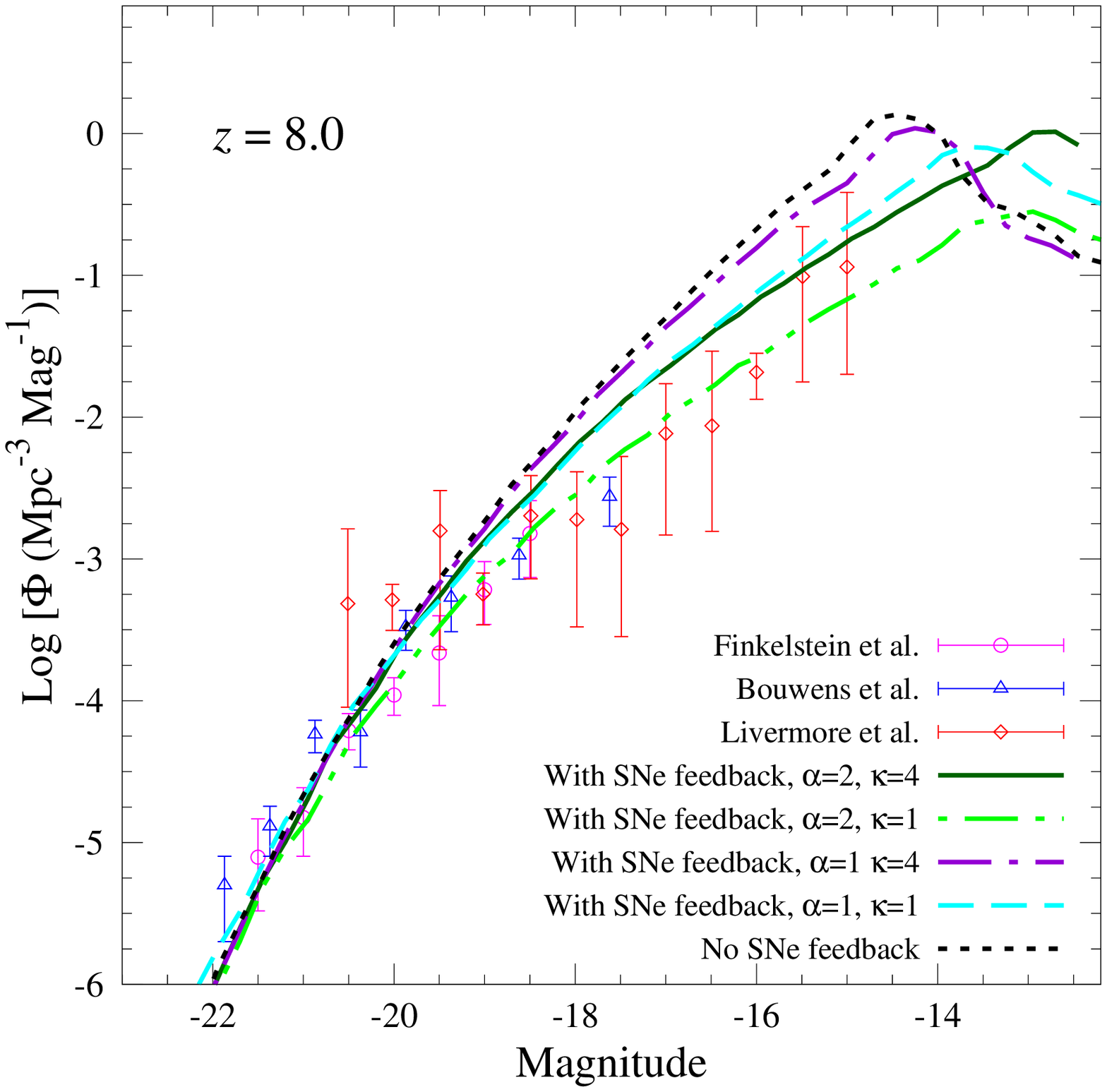,width=6.0cm,angle=-90.}%
\epsfig{figure=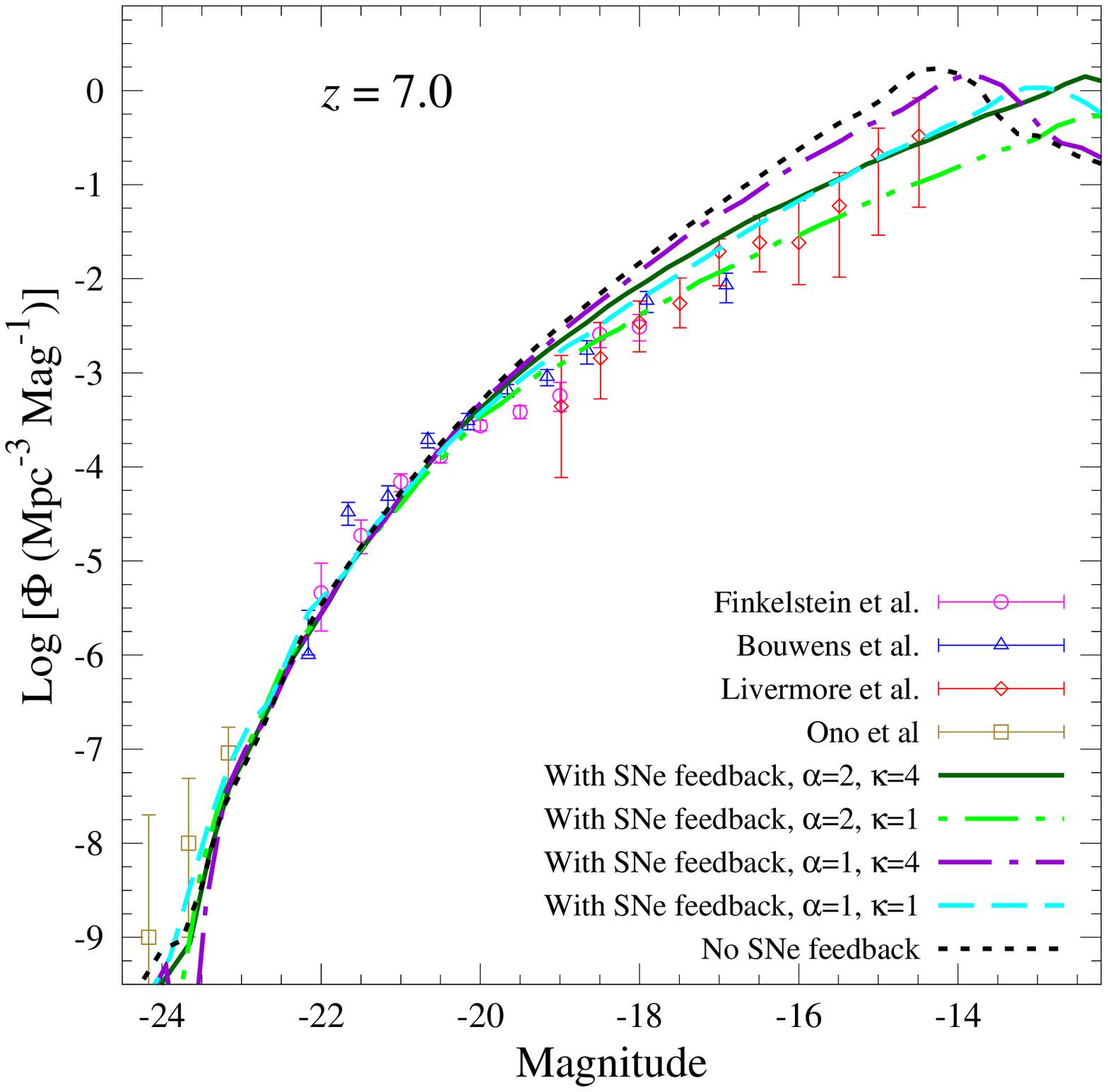,width=6.0cm,angle=-90.}%
\epsfig{figure=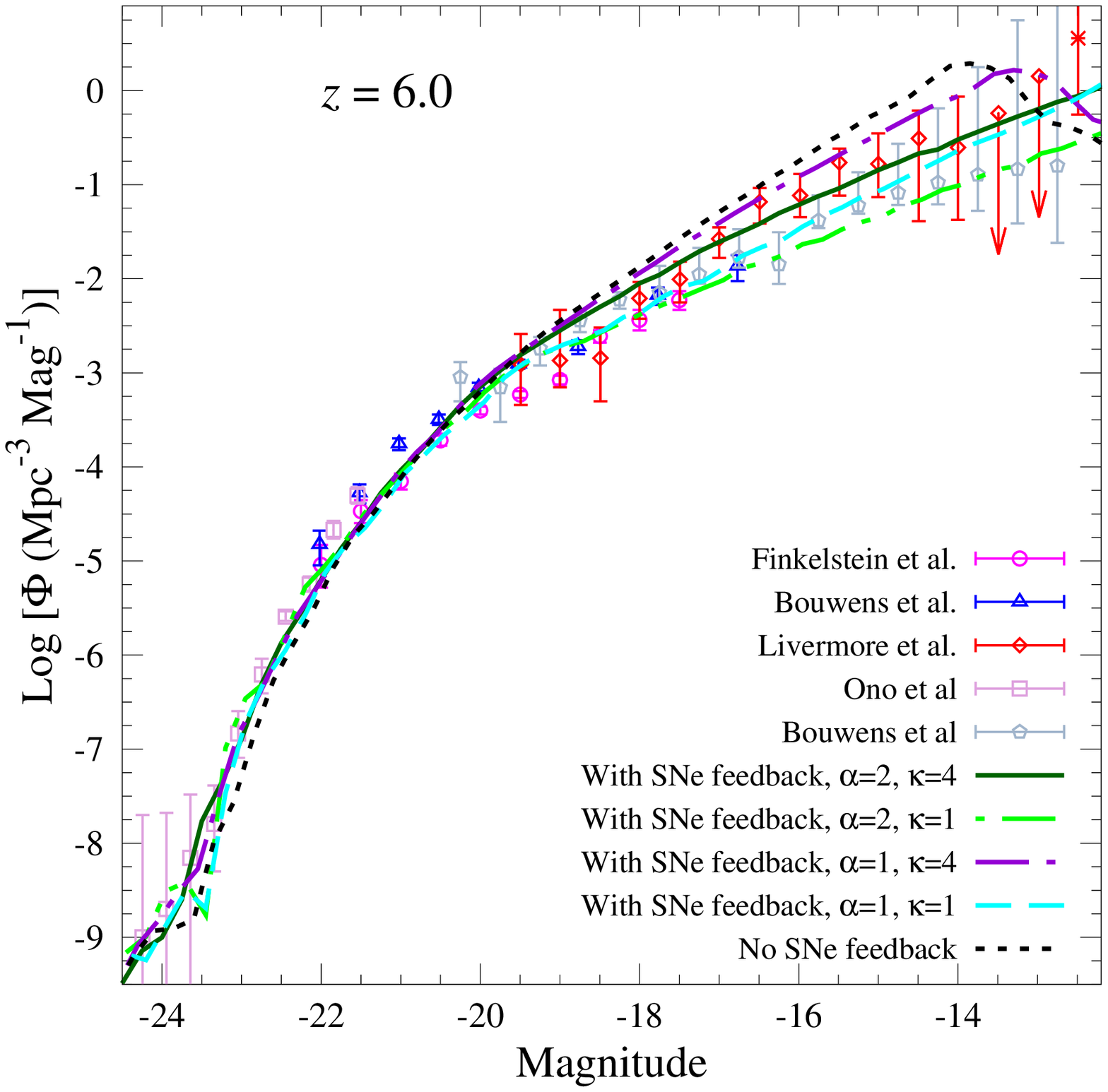,width=6.0cm,angle=-90.}%
}
\caption[]{UV luminosity functions of LBGs at $z=8$, 7 and 6 (from left to right panel) as predicted by
our models along with the observational data. The data are taken from
\citet{2017ApJ...835..113L} (open diamond), \citet{2015ApJ...803...34B}
(open triangles), \citet{2015ApJ...810...71F} (open circles), \citet{2018PASJ...70S..10O}
(open squares) and \citet{2017ApJ...843..129B} (open pentagons).
The solid and dot dot dashed curves show the prediction of our SNe feedback models with
$\eta_w=(v_c/100~{\rm km/s})^{-2}$ for $\kappa=4$ and 1 respectively.
The dotted-dashed and dashed curves are for SNe feedback models with
$\eta_w=(v_c/100~{\rm km/s})^{-1}$ for $\kappa=4$ and 1 respectively.
Model predictions for without SNe feedback are shown by the dotted lines.}
\label{fig_lf_z6-8}
\end{figure*}

We begin by comparing our model predictions
with observations at $z=8$, 7 and 6
(see Fig.~\ref{fig_lf_z6-8}) when HI in the universe is expected to transit from neutral to ionised state
(i.e. going through the final stages of HI reionisation). 
For each redshift we show predicted luminosity functions from five different models.
The dotted curves represent the luminosity functions obtained
for the model without the supernova feedback in
star formation ({\it i.e.} $\eta_w=0$). However, this model includes
radiative feedback. It is clear from the figure that in all three
redshifts, such a model over predicts the observed number count of
galaxies in low luminosity end, M$_{\rm UV}\gtrsim -19$.
Thus the radiative feedback alone is not enough
to reproduce the observed shape of the luminosity function at low
luminosity end.
Further note that given
the small value of $\tau_e$ from CMBR data one would expect to have
less than 10\% of the universe to be ionised at $z=8$ and indeed
we verify that in our self consistent reionisation models as well. Hence, at $z=8$
the radiative feedback would be expected to affect the luminosity function by
at most 10\%. Thus the small suppression due to reionisation
feedback is not enough to explain the shape of observed
luminosity functions at low luminosity end in these redshifts and this is almost insensitive to
the model of ionisation feedback that one assumes. Further, note that there is a turn over seen in  the predicted
luminosity functions at M$_{\rm UV}\sim -15$ for this model. This turn over arises
due to the atomic cooling cut off in star formation
of halos with virial temperature less than $10^4$~K and comes within the
observed luminosity range
at $z=6$. It is also independent of reionisation model. In absence of such turn
over in the observational data, we conclude
that only radiative feedback alone can not explain both the shape and extent of UV
luminosity functions. Therefore, we need additional feedback other than the radiative
feedback to explain the observation.

Indeed, we show that the SNe feedback model leads to correct shape of the UV luminosity functions.
In Fig.~\ref{fig_lf_z6-8}, we show luminosity functions predicted by energy
driven supernova feedback models ({\it i.e.} $\alpha=2$) with $\kappa=4$ (solid curves) and $\kappa=1$
(dot-dot-dashed curves). We also show model predictions for momentum driven 
supernova feedback models ({\it i.e.} $\alpha=1$) with $\kappa=4$ (dotted dashed curves) and $\kappa=1$
(dashed curves). 
With such feedbacks, the atomic cooling turn over got shifted to even lower luminosity
as the same galaxy would have become fainter due to the feedback. This has
also reduced the number counts of galaxies with M$_{\rm UV}>-19$, required by the observations.
Thus we can say that the new faint end luminosity
data clearly shows the SNe feedback in action in the redshift range $z=6-8$.

Further, the faint end luminosity functions can in principle be used to constrain
the nature of supernova feedback and duration of star formation activity.
We see from Fig.~\ref{fig_lf_z6-8} that each models predict
distinct luminosity functions for  M$_{\rm UV}>-19$. Both the atomic cooling
turn over and amplitude of luminosity functions are different for different
models. Although, the observations at $z=7$ and 8 are not yet accurate enough to distinguish
between models, the redshift $z=6$  observation can. The atomic cooling turnover
for $\kappa=4$ and $\alpha=1$ model arises at brighter luminosity compare to $\alpha=2$ model.
At $z=6$, such turn over is expected to appear at  M$_{\rm UV}\sim -13$ for
$\alpha=1$ model.

Note that there are discrepancies of $z=6$ observed luminosity functions
as analysed by different groups, especially in the uncertainty of
the very faint end of the luminosity function
\citep[see][for a detailed discussion]{2017ApJ...843..129B}. In Fig.~\ref{fig_lf_z6-8}, we show
results from \citet{2017ApJ...843..129B} (open pentagons)
as well as from \citet{2017ApJ...835..113L}(open diamonds).
The discrepancy is clear in the figure. While the data from \citet{2017ApJ...835..113L}
prefers $\kappa=4$ model, the \citet{2017ApJ...843..129B} data 
favours $\kappa=1$ model (along with $\alpha=2$). Thus once the observational debate is settled
we will be able to constrain the duration of star formation activities in
those high redshift galaxies. In passing we mention that lower redshift
observation favours a $\kappa=4$ model (see below). Note that with such values of $\kappa$,
a galaxy is expected to have active star formation for a longer period
and thus one
would not expect to see the passively
star forming galaxies at higher redshifts. We will discuss this in detail later.

Further, it is clear from the Fig.~\ref{fig_lf_z6-8} that our models provide
good fit to the observational data of \citet{2018PASJ...70S..10O} in the high luminosity range (i.e. $-24\le M_{\rm UV}<-22$)
where AGN feedback is important in deciding the shape of the luminosity
functions. Thus, the AGN feedback model that we adopted here is adequate to explain the
observational data. We return to discussions on  the importance of
AGN feedback at different redshifts in Section~\ref{sec_pr}.

Therefore, we see that our star formation model including
energy driven outflow feedback by SNe in star formation (i.e. $\eta_w \propto v_c^{-2}$)
with $\kappa = 4$ explains shape and the redshift evolution of observed UV
luminosity function of galaxies over a wide luminosity range $-24\le M_{\rm UV}\le -12.5$
from $z=8$ to $z=6$. We call this model as our fiducial model
and will show next that this model can also reproduce the observed
luminosity functions at lower redshifts.

\subsection {Post reionisation period: $z=5,$ $4,$ $3$}
\label{sec_pr}
We now turn into the post reionisation era as the quasar spectrum
confirms the end of reionisation by $z\sim 6$. 
Note that for these redshifts
we do not have very faint end luminosity functions from the gravitational
lensing measurements
to put tight constraints on the nature of the star formation.
Here we just show that our fiducial model provides good fit to the available data. Future
observations of the faint galaxies would confirm our predictions.

\begin{figure*}
\centerline{
\epsfig{figure=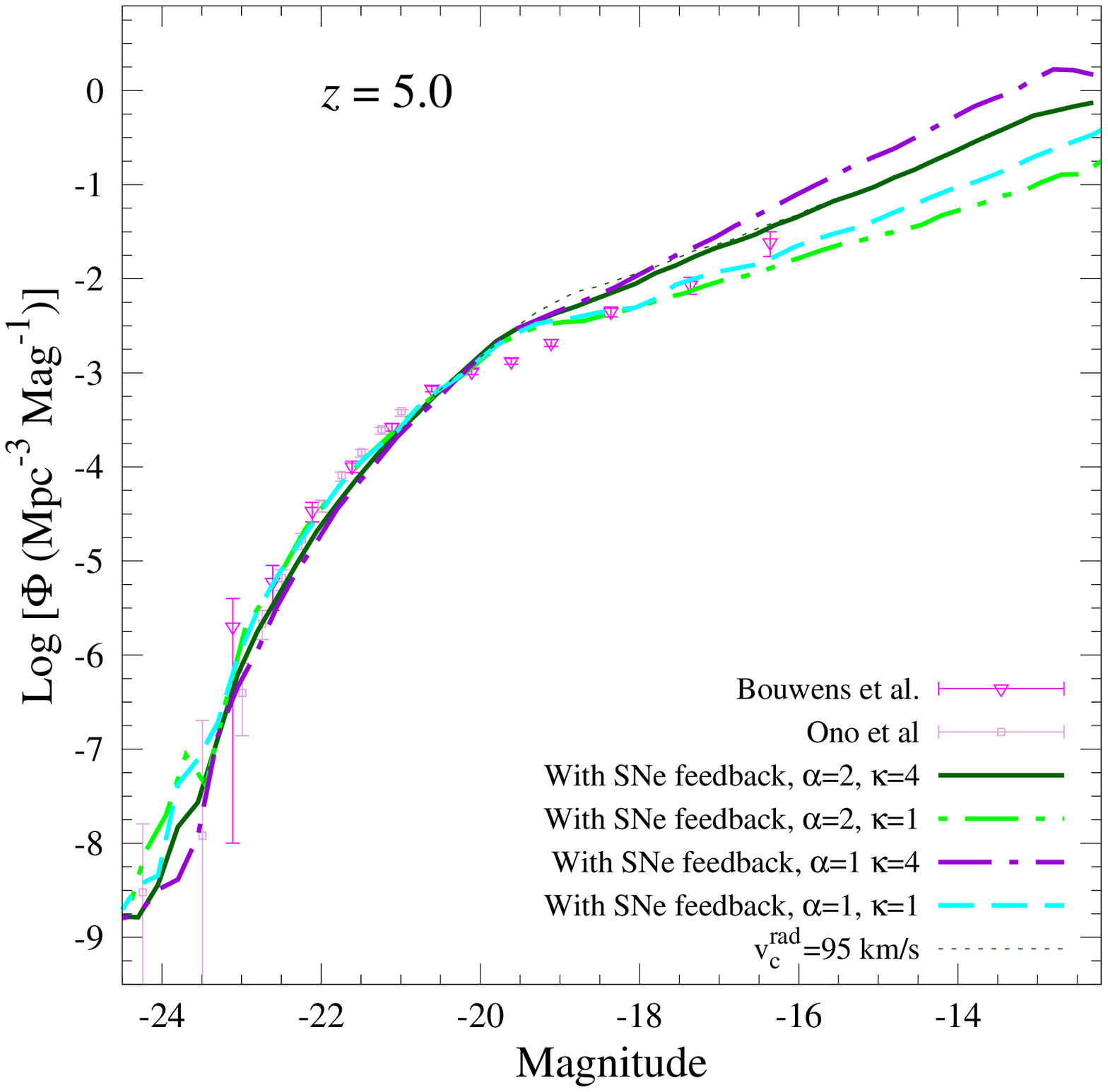,width=6.0cm,angle=-90.}%
\epsfig{figure=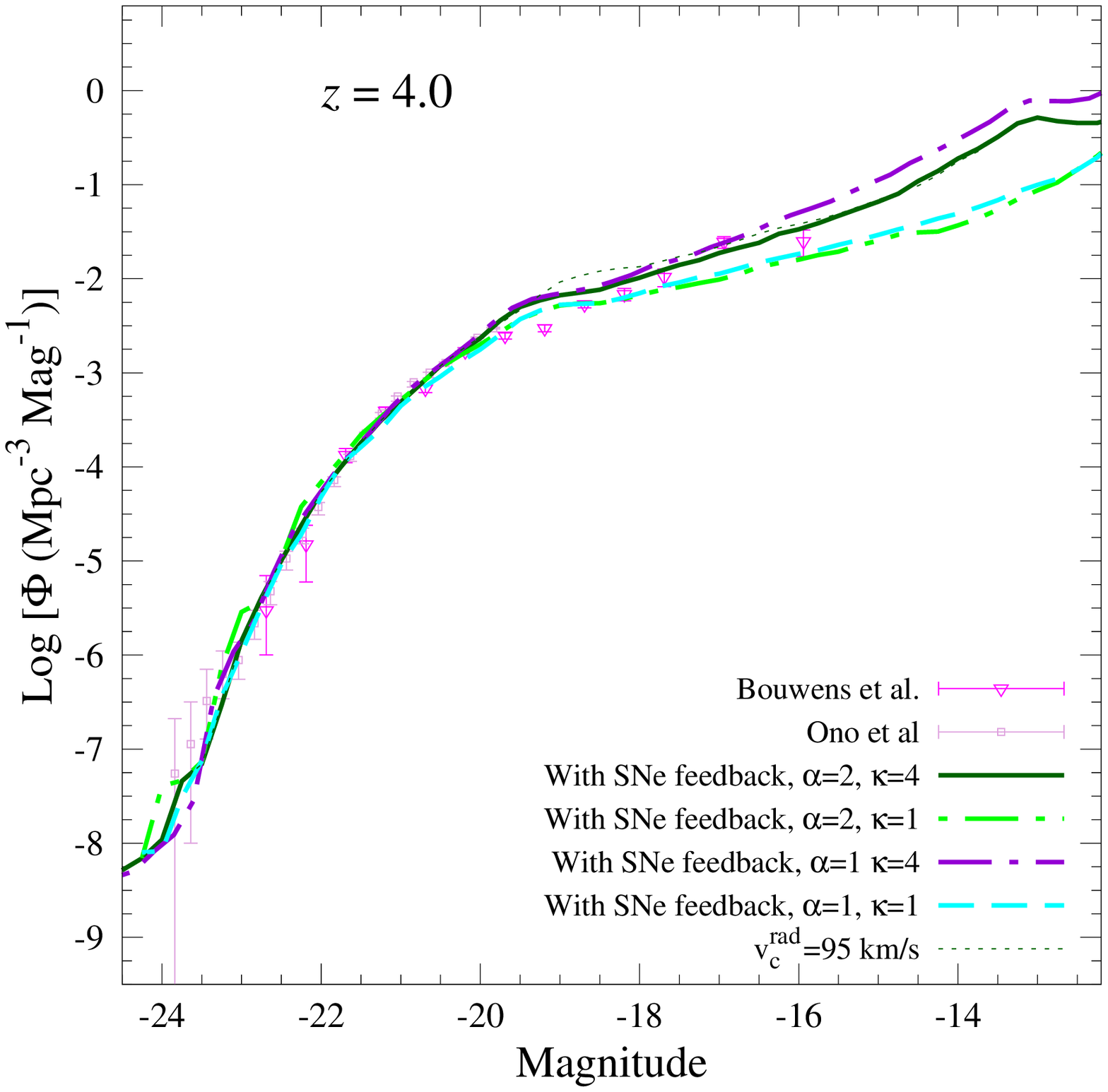,width=6.0cm,angle=-90.}%
\epsfig{figure=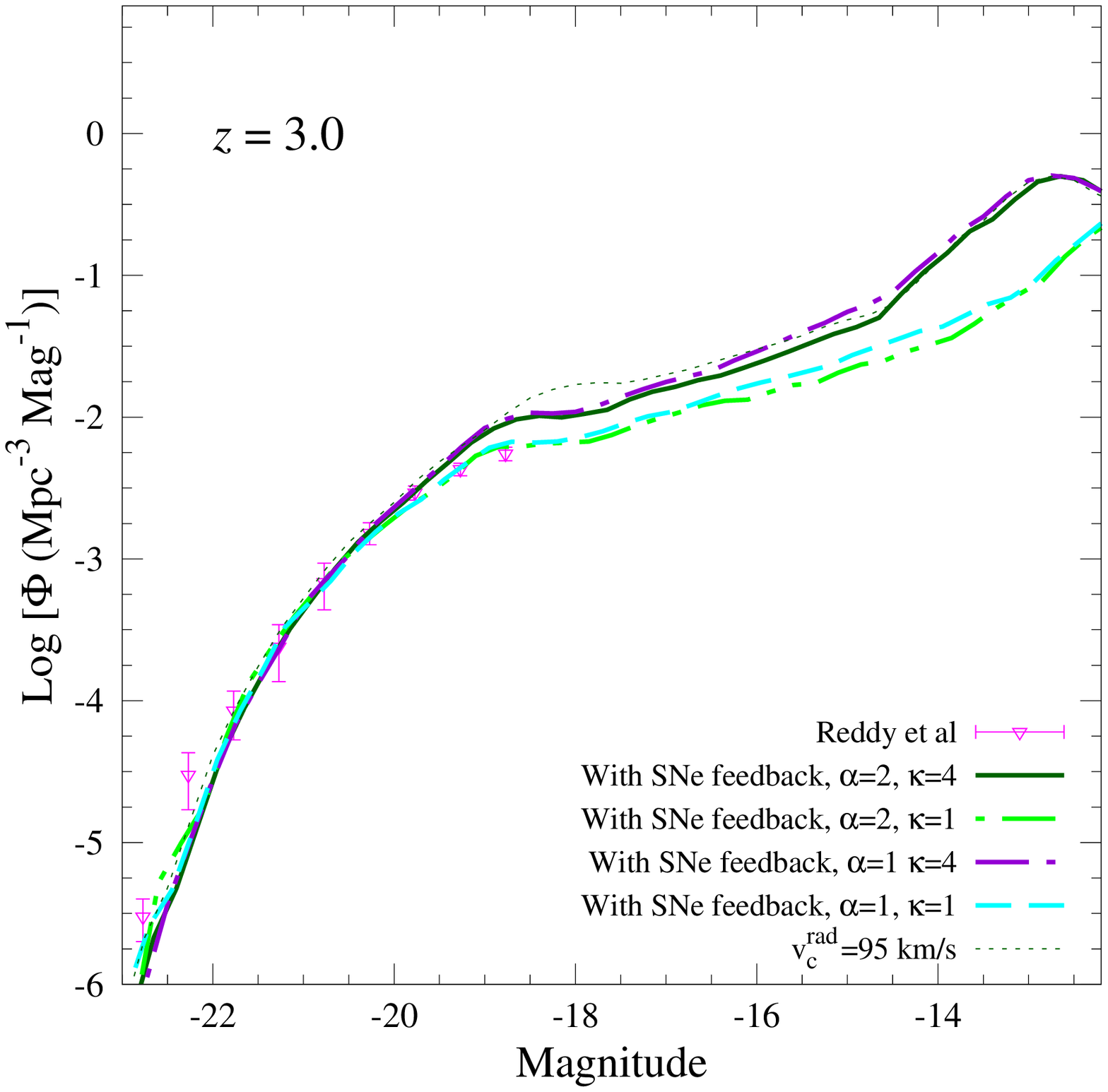,width=6.0cm,angle=-90.}%
}
\caption[]{UV luminosity functions of LBGs at $z=5$, 4 and 3 as predicted by
our models along with the observational data. The data are taken from
\citet{2015ApJ...803...34B}, \citet{2018PASJ...70S..10O} and \citet{2009ApJ...692..778R}.
The solid and dot dot dashed curves show the prediction of our SNe feedback models with
$\eta_w=(v_c/100~{\rm km/s})^{-2}$ for $\kappa=4$ and 1 respectively.
The dotted-dashed and dashed curves are for SNe feedback models with
$\eta_w=(v_c/100~{\rm km/s})^{-1}$ for $\kappa=4$ and 1 respectively.
The dotted lines show the effect
of radiative feedback being only extend upto the halo with circular
velocity $v_c^{rad}=95$~km/s.}
\label{fig_lf_z3-5}
\end{figure*}

\begin{figure*}
\centerline{
\epsfig{figure=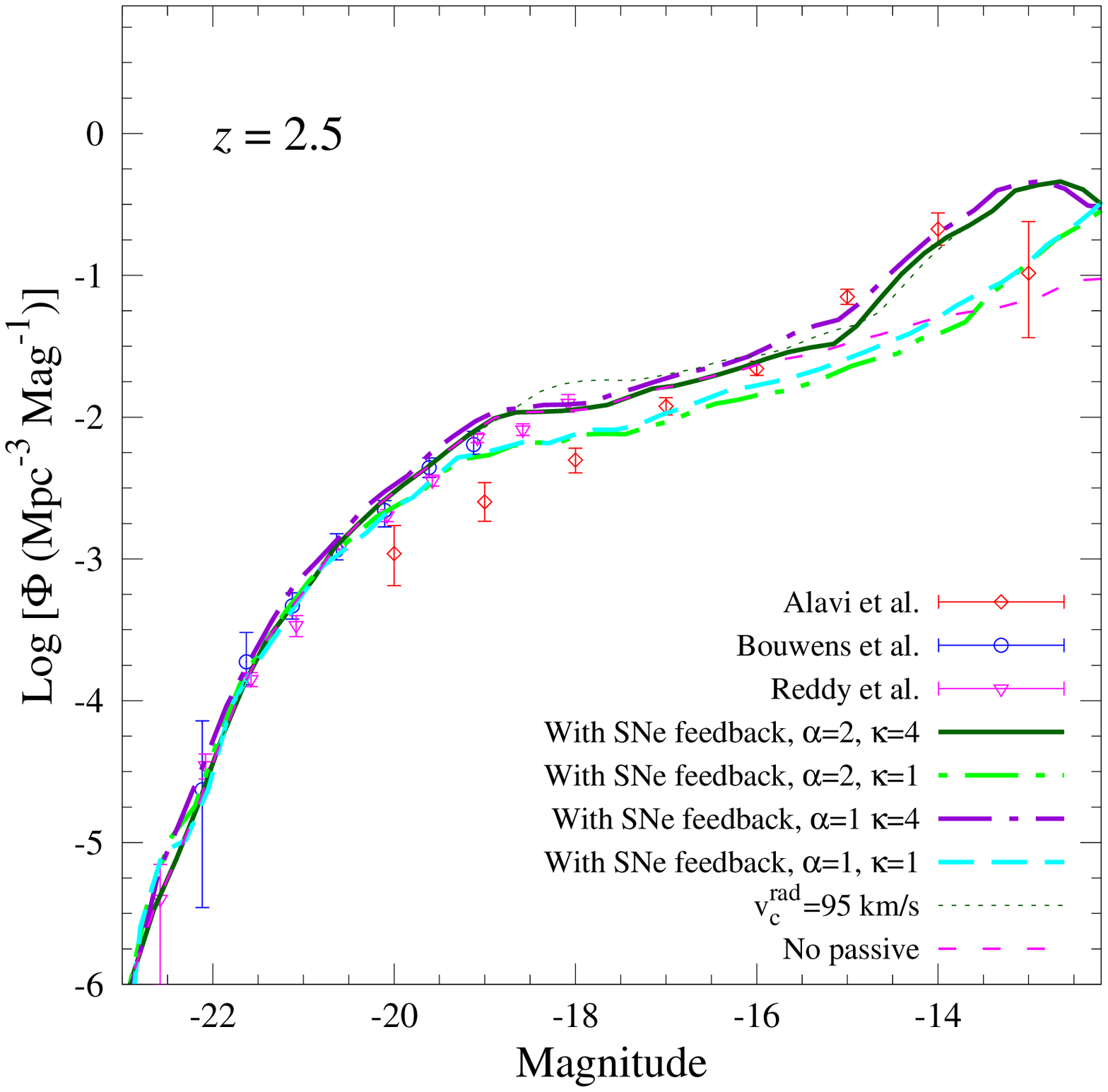,width=6.0cm,angle=-90.}%
\epsfig{figure=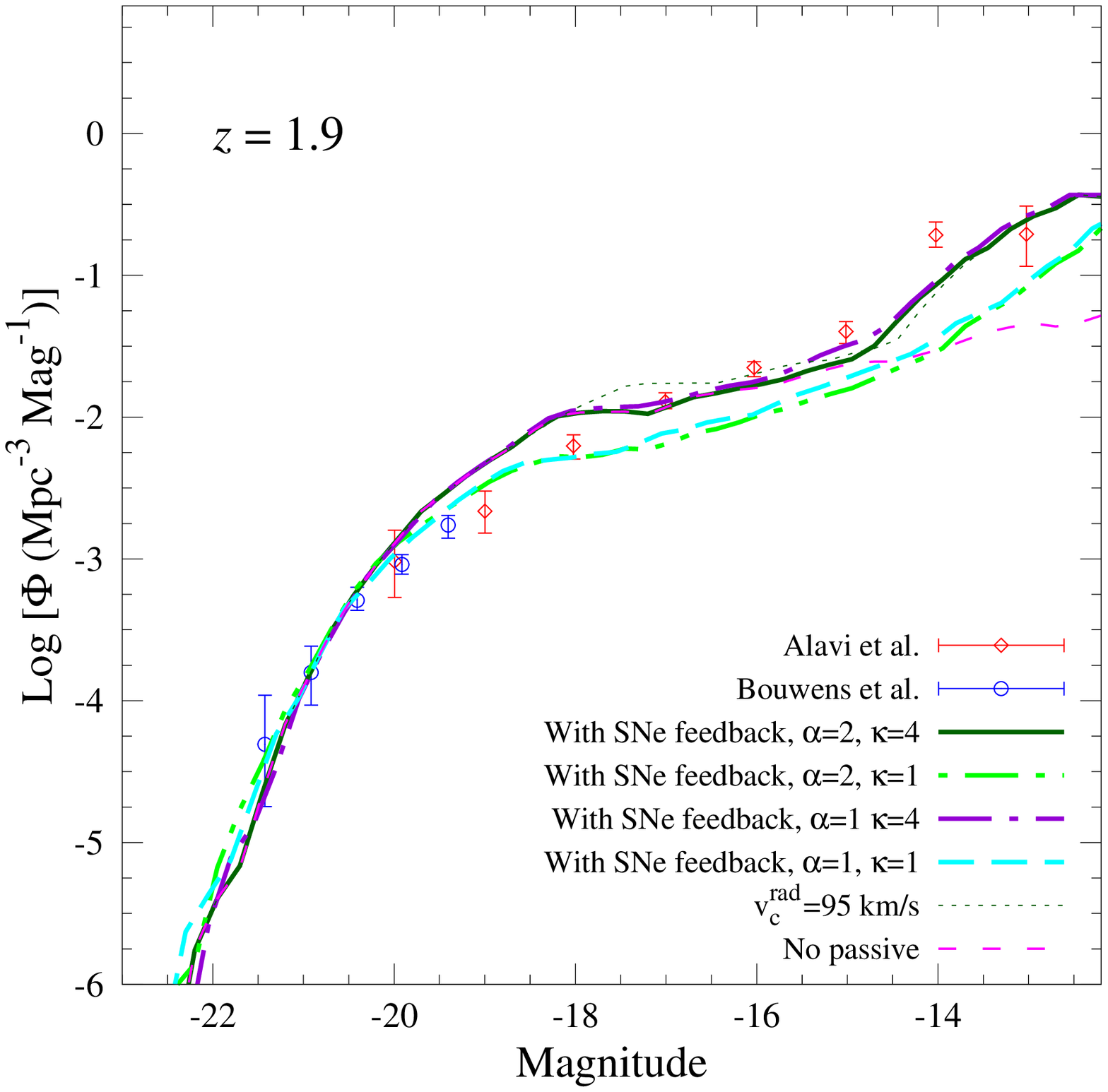,width=6.0cm,angle=-90.}%
\epsfig{figure=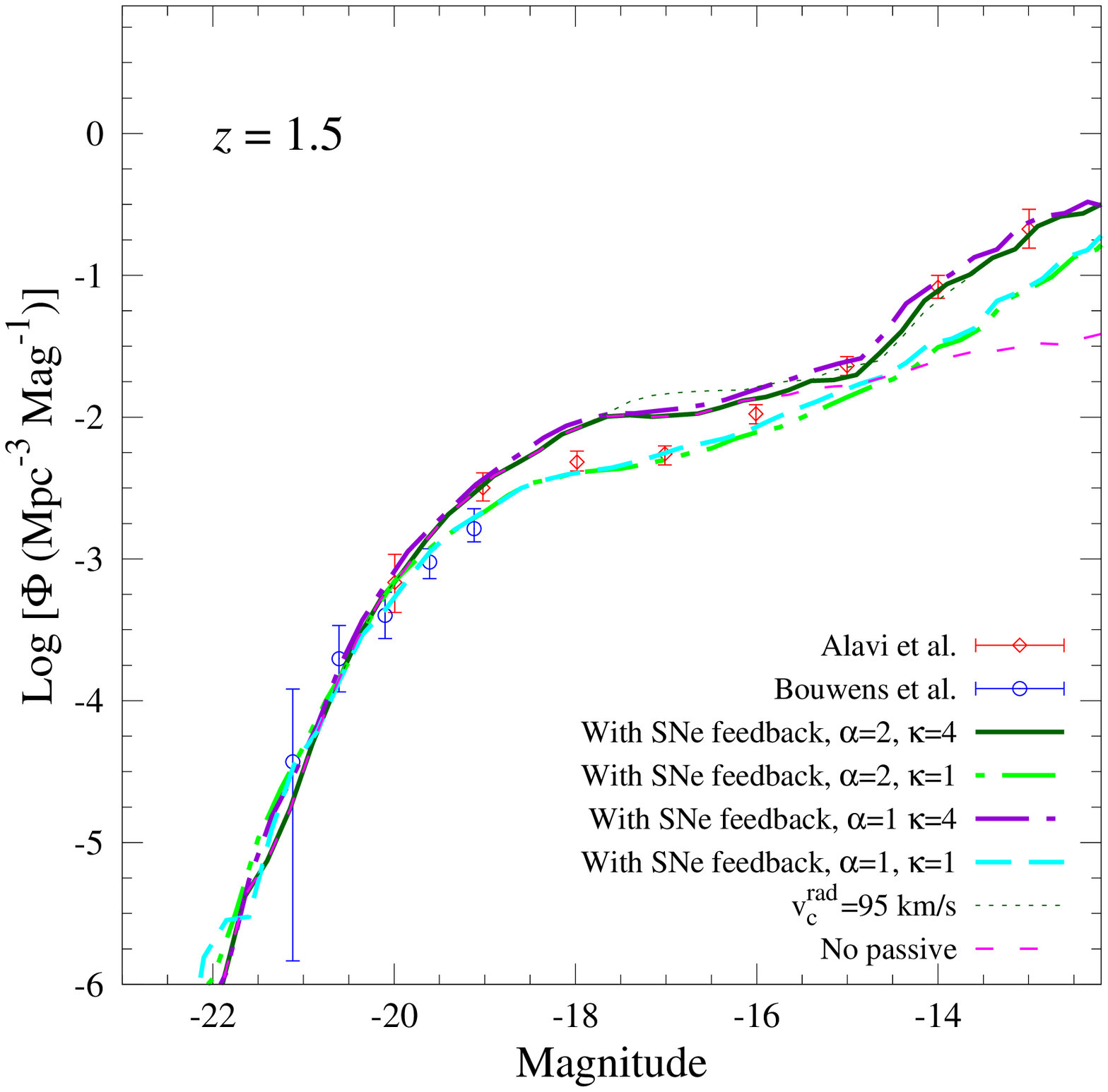,width=6.0cm,angle=-90.}%
}
\caption[]{UV luminosity functions of LBGs at $z=2.5$, 1.9 and 1.5 as predicted by
our models along with the observational data. The data are taken from \citet{2009ApJ...692..778R} (open triangles),
\citet{2015ApJ...803...34B} (open circles) and \citet{2016ApJ...832...56A} (open diamonds).
The solid and dot dot dashed curves show the prediction of our SNe feedback models with
$\eta_w=(v_c/100~{\rm km/s})^{-2}$ for $\kappa=4$ and 1 respectively.
The dotted-dashed and dashed curves are for SNe feedback models with
$\eta_w=(v_c/100~{\rm km/s})^{-1}$ for $\kappa=4$ and 1 respectively.
The dashed lines are for models without the passive mode of star formation
(and other parameters are same as the fiducial model) . The dotted
curves are for models with change in radiative feedback
cut of circular velocity to $v_c^{rad}=95$~km/s.}
\label{fig_lf_z1-3}
\end{figure*}

Fig~\ref{fig_lf_z3-5} shows our model predictions and compare them with the observed data. 
It is evident from these figures that difference between the model predictions are
larger at M$_{\rm UV}\ge -17$ where the observations are sparse at present. Interestingly
none of our models predict turn over of the faint end slope 
of the luminosity functions down to M$_{\rm UV}\sim -13$ in these
redshifts as well.
The predictions of SNe feedback models with  $\eta_w \propto v_c^{-2}$
and $\kappa=4$ (solid curves)
match very well with the observed data points. Moreover, at
$z=3$ they predict even an increase in the faint end slope at
M$_{\rm UV}\gtrsim -15$. We will show later that this enhancement
is coming from the galaxies that are currently evolving with the
passive mode of star formation (i.e. Fig.~\ref{fig_mass}) and indeed
been observed at even lower redshifts (see Fig.~\ref{fig_lf_z1-3}).
Thus we predict that the next generation telescopes which can observe very faint
galaxies will see this extended faint end luminosity
functions upto M$_{\rm UV} \sim -13$ at $z=5, ~4$ and an
enhancement in the number counts
of faint galaxies with M$_{\rm UV} \gtrsim -15$ at $z=3$ if passive galaxy population
contributes to the observed luminosity functions at these redshifts as they do at
low-$z$.

Further, we can infer from  Fig.~\ref{fig_lf_z3-5} that
observations at these redshifts can also constrain the halo mass upto which
the radiative feedback is effective. The break seen in the observed luminosity
functions at M$_{\rm UV} \sim -19$ is created by the radiative feedback as
predicted by our models that assume
such feedback is operating upto haloes of circular velocity
110~km/s. Models that assume a radiative feedback in halos with
virial velocity less than 95~km/s (dotted curves in Fig.~\ref{fig_lf_z3-5})
clearly over predict the observed luminosity functions.
Thus even more massive systems are prone to the ionisation feedback
compared to what has been found in numerical simulations
\citep{1996ApJ...465..608T}. Similar results were
obtained by \citet{2014MNRAS.443.3341J} as well.

\begin{figure}
\centerline{
\epsfig{figure=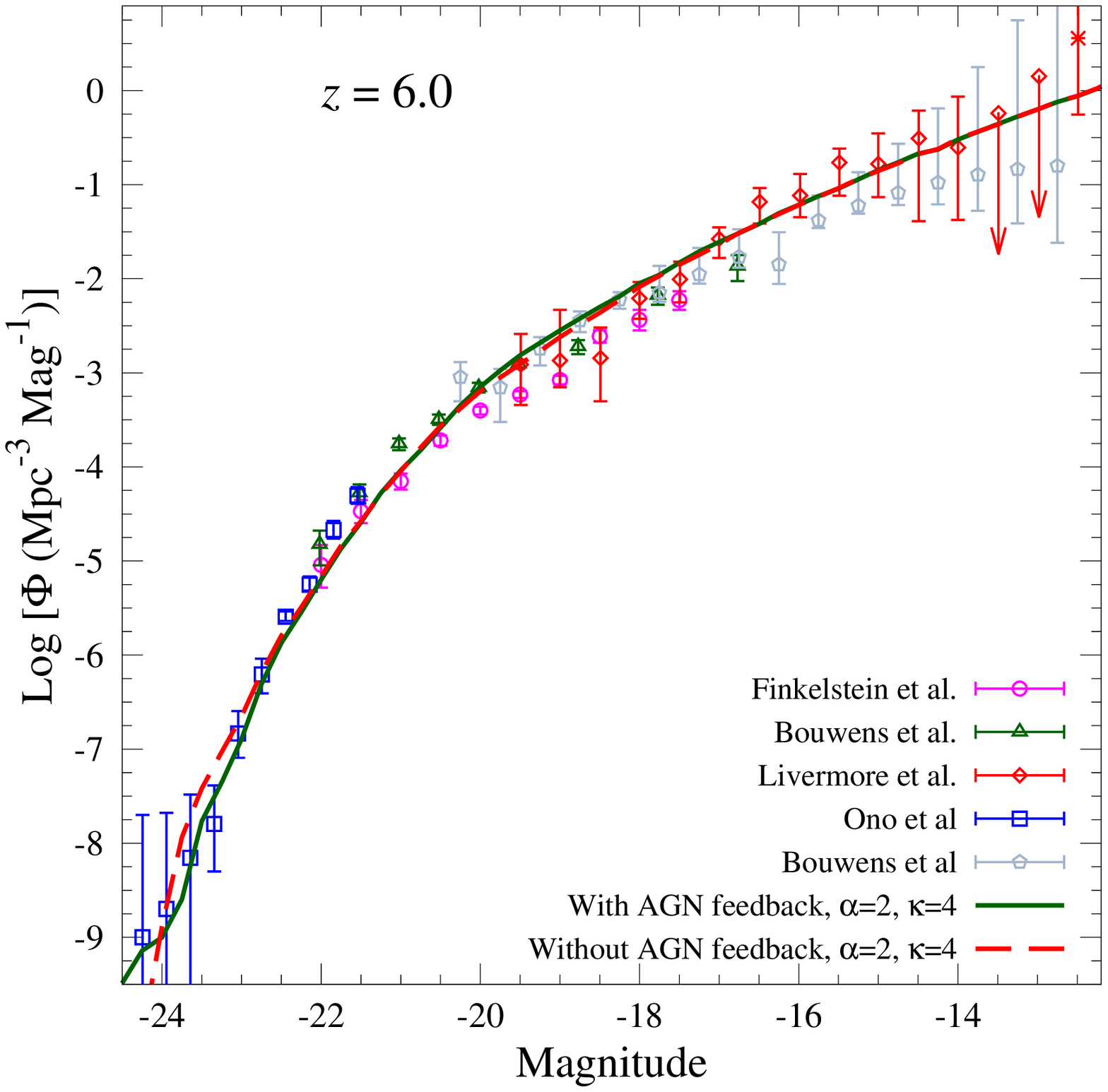,width=6.5cm,angle=-90.}%
}
\centerline{
\epsfig{figure=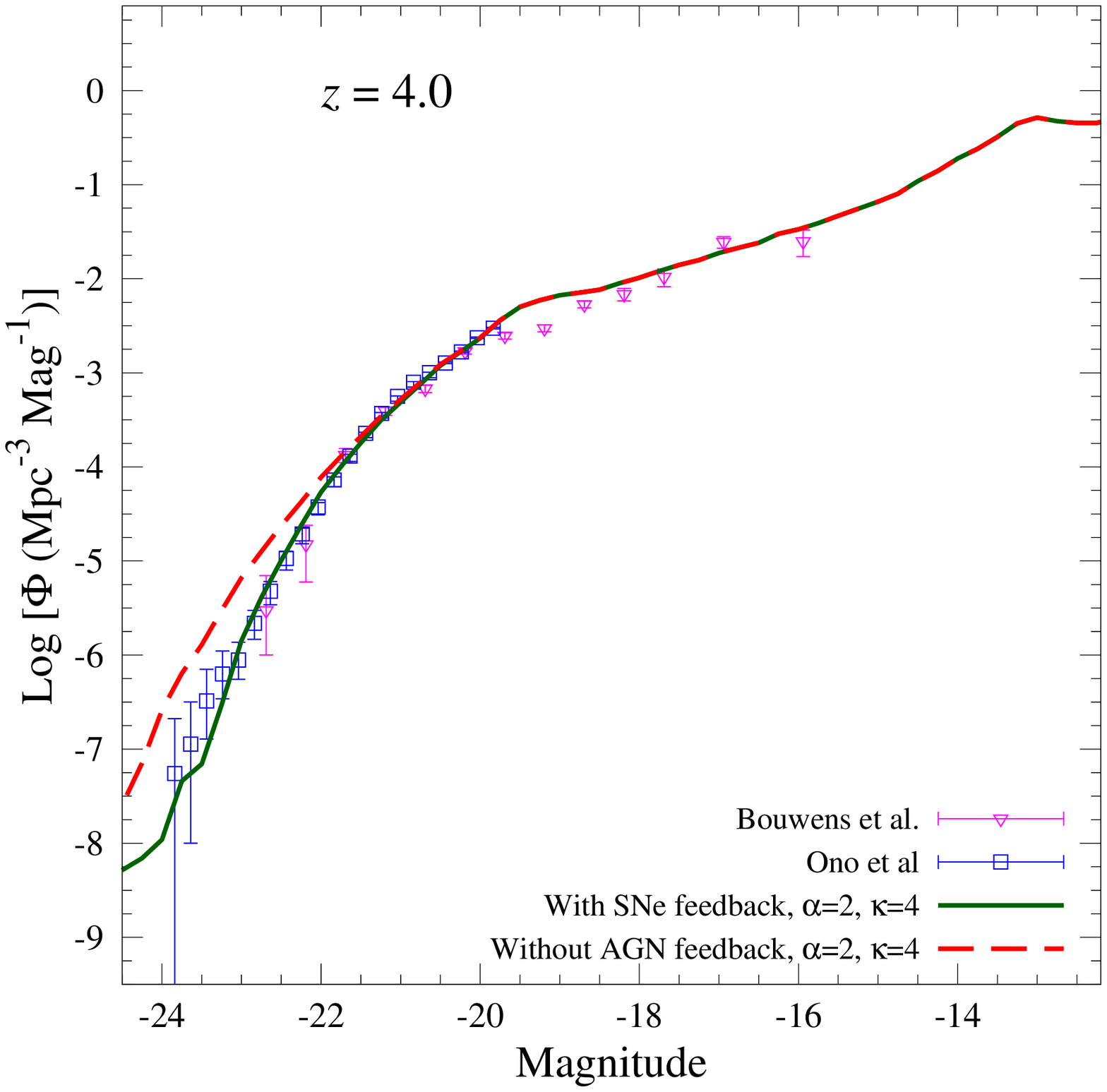,width=6.5cm,angle=-90.}%
}
\caption[]{The effect of AGN feedback on the luminosity functions at $z=4$ (bottom panel) and $z=6$ (top panel). The solid lines
show luminosity functions obtained from our fiducial model whereas the dashed lines show luminosity functions
as predicted by a model that does not consider AGN feedback.
}
\label{fig_AGN}
\end{figure}

Finally we show that the observed data points at high luminosity end (i.e. M$_{\rm UV} < -22$)
can indeed show the signature of AGN feedback. In the lower panel of Fig.~\ref{fig_AGN} we have plotted the observed luminosity
function at $z=4$ along with predictions from two models: (i) with AGN feedback (solid line) (ii) without
AGN feedback (dashed line). It is clear that the model which does not consider the AGN feedback
over predicts the number counts of galaxies in the luminosity range M$_{\rm UV} < -22$. Only the
model considering AGN feedback explains the observed number counts of galaxies.
In passing we note that the difference between the predictions of the models with and
without AGN feedback
increases with decreasing redshifts. In Fig.~\ref{fig_AGN}  the requirement of AGN feedback
is more clearly demonstrated for $z\sim 4$ compared to that  for $z\sim 6$.
At $z=6$ the effect 
of AGN feedback is small and lies within the uncertainty of presently available observations
(top panel of Fig.~\ref{fig_AGN}).

\subsection {Late universe: $z=2.5,$ $1.9,$ $1.5$}

Now we discuss luminosity functions at even lower redshifts ($z=2.5$, 1.9 and 1.5)
 where \citet{2016ApJ...832...56A} provide measurements
upto very faint end of M$_{\rm UV}=-13$. In these redshifts also the observed UV
luminosity functions of galaxies do not show any turn over upto
a magnitude limits of M$_{\rm UV}=-13$. Moreover, they show an increase
in the slope of the luminosity functions at M$_{\rm UV}\gtrsim -15$.
Thus it is interesting to see if the star formation models that are successful
in describing the high redshift universe also able to explain the low redshifts
counterparts. 

Indeed we see in Fig.~\ref{fig_lf_z1-3} that the same fiducial model
of star formation provides good fit to the observational data for
all three redshift bins
and in the entire luminosity range (the solid lines in Fig.~\ref{fig_lf_z1-3}).
Like the other higher redshift bins, prolonged star formation mode ({\it i.e.} $\kappa=4$)
is favoured compared to a burst mode ({\it i.e.} $\kappa=1$)
at these epochs as well. However, we can not constrain the nature
of supernova feedback (i.e the choice of $\alpha$) in these redshifts
as luminosity functions predicted by
models with different $\alpha$ for a given $\kappa$ differ very little.

The most striking feature
at these redshifts is the very faint end luminosity function
(i.e. magnitude $\gtrsim -15$) where there is 
an enhancement in the observed number density of
faint galaxies. Some of our models indeed reproduce such enhancement.
A detailed investigation reveals that such an enhancement in our
models is mainly due to
the older generations of galaxies that are exhibiting the slow passive mode
of star formations.
This can be understood as follows. In Fig.~\ref{fig_lf_z1-3} we also show the
model predictions that do not consider this passive star formation by the
magenta dashed lines. Clearly in such models the luminosity functions
are much flatter than the observed ones and under predict
the number counts of galaxies of faint magnitude M$_{\rm UV}\gtrsim -15$
in all three redshift bins. Only models that consider
the passive slow mode of star formation would be able to explain the
number counts of faint galaxies at these redshifts. Note that the faint end
turn over as predicted by our models now occurs at M$_{\rm UV}\gtrsim -12$ in these redshifts
and future observations extending to such low luminosities are likely to detect it.

\begin{figure*}
\centerline{
\epsfig{figure=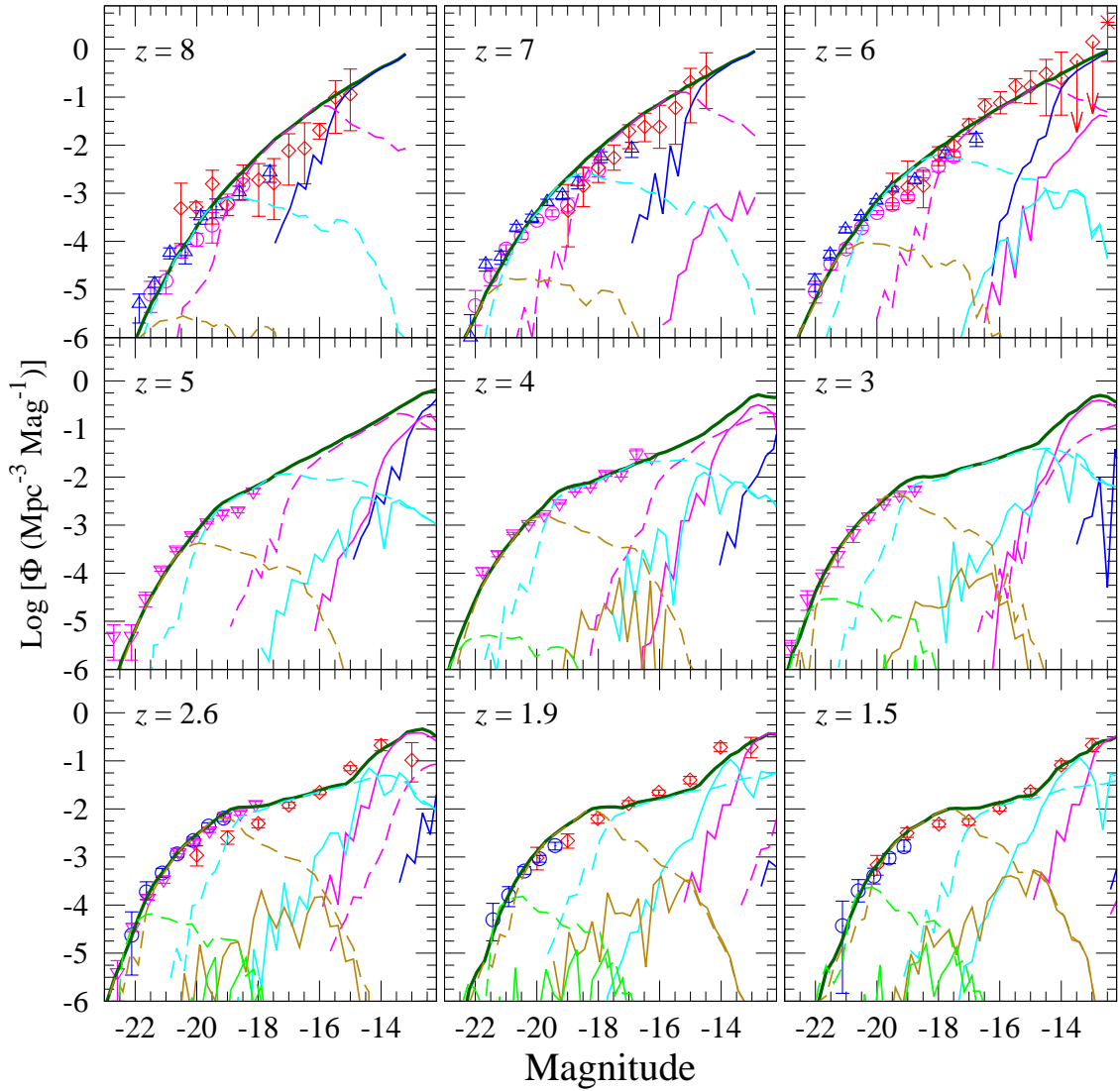,width=15.0cm,angle=-90.}}
\caption[]{Luminosity functions at different redshifts. The solid dark-green lines
show predicted luminosity functions by our fiducial SNe feedback models. We show the break up of active and passive
mode of star formation by dashed and solid lines respectively. They
are further broken up to show contribution from different galaxy mass
ranges shown by different colour curves: $10^{8}-10^{9}$~M$_\odot$ (blue),
$10^{9}-10^{10}$~M$_\odot$ (magenta), $10^{10}-10^{11}$~M$_\odot$ (cyan),
$10^{11}-10^{12}$~M$_\odot$ (golden), $10^{12}-10^{13}$~M$_\odot$ (green).
}
\label{fig_mass}
\end{figure*}

To show the effect of this passive mode of star formation in different redshifts
we have plotted the break up of luminosity functions contributed by the
galaxies with ongoing active star formation and passively star forming galaxies in Fig.~\ref{fig_mass}.
In all redshift bins we also show contributions arising from galaxies with
different mass ranges. The contributions by the active star forming
galaxies are shown by different color dotted lines whereas the
same for the passively star forming galaxies are shown by the solid
curves. From the figure is it clear that in high redshifts i.e.
$z\ge6$ the faint end luminosity functions are still contributed
by the active star forming galaxies with masses in the range
$10^{9}~{\rm M}_\odot \le M \le 10^{12}~{\rm M}_\odot$. The universe
is not old enough at these redshifts to have evolved galaxies that
are going through the passive slow mode of star formation. Further,
the new observations by \citet{2017ApJ...835..113L} using the
gravitational lensing
magnifications are detecting dwarf galaxies with masses
$10^9~{\rm M}_\odot \le M \le 10^{10}~{\rm M}_\odot$ at $z\ge6$. These
galaxies contribute most to the UV photon budget for the
reionisation. In redshift
range $ 5 \ge z \ge 3 $ the observations are upto M$_{\rm UV} \lesssim -18$
and only the active star forming galaxies of masses
$10^{10}~{\rm M}_\odot \le M \le 10^{12}~{\rm M}_\odot$
contribute to the observed galaxies. Future observations extending
to fainter magnitude are likely to detect these passive star forming galaxies
at those redshifts.

Only in redshift range $1.5 \le z \le 2.5$ where the observed luminosity
functions extend to as faint as M$_{\rm UV}=-12.5$ we see the contribution
from the passively star forming galaxies of masses
$10^{9}~{\rm M}_\odot \le M \le 10^{11}~{\rm M}_\odot$ with magnitude
M$_{\rm UV} \gtrsim -15$. The higher luminosity range is contributed
by the actively star forming galaxies with mass range
$10^{10}~{\rm M}_\odot \le M \le 10^{13}~{\rm M}_\odot$ at
$1.5 \le z \le 2.5$. Thus we can say that the gravitational
lensing measurements enable us to probe
small mass galaxies that are evolving passively at these redshifts. These are
the counterparts of local red sequence galaxies.

\subsection{Dependence on model parameter}
\begin{figure*}
\centerline{
\epsfig{figure=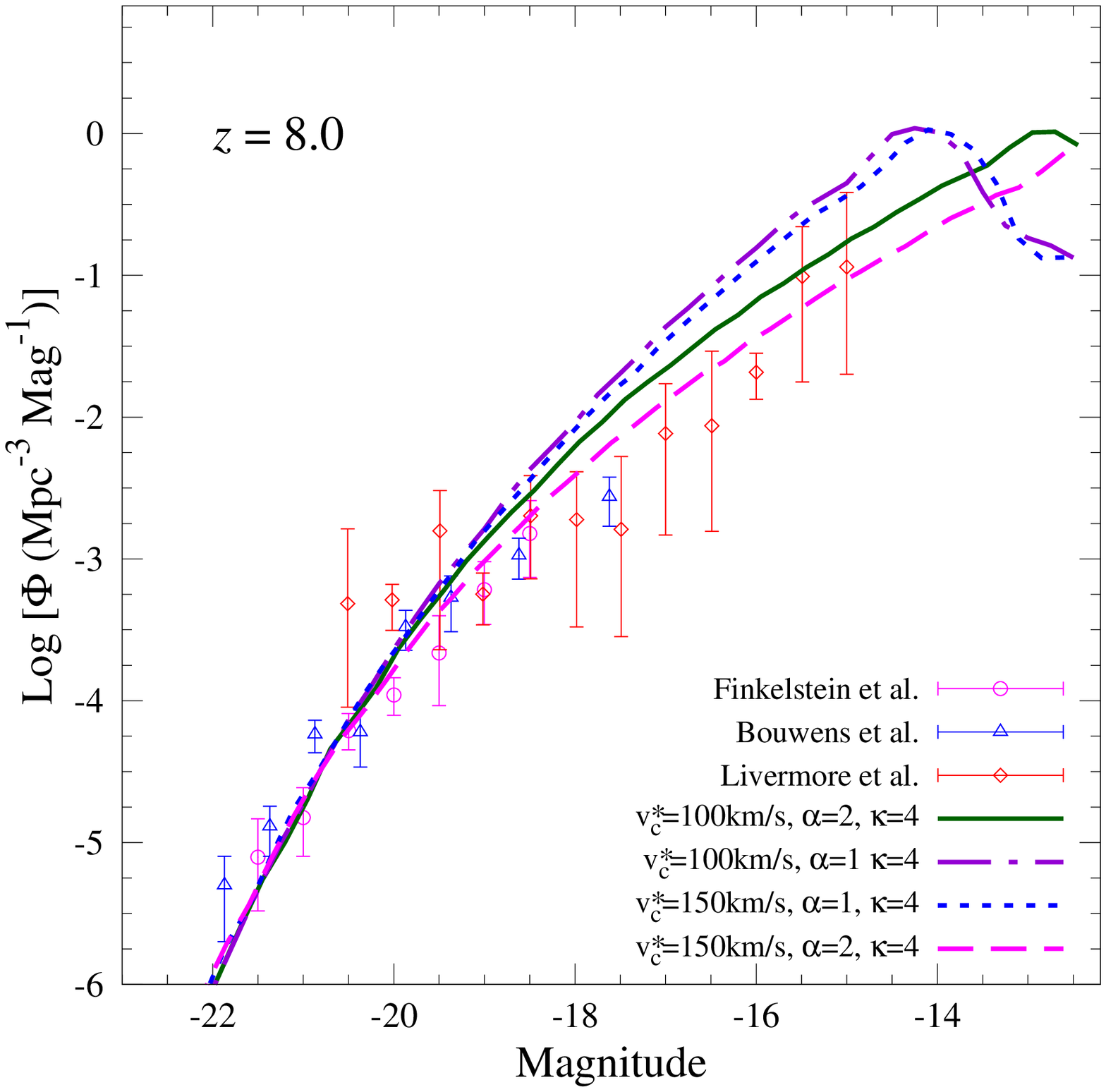,width=5.5cm,angle=-90.}%
\epsfig{figure=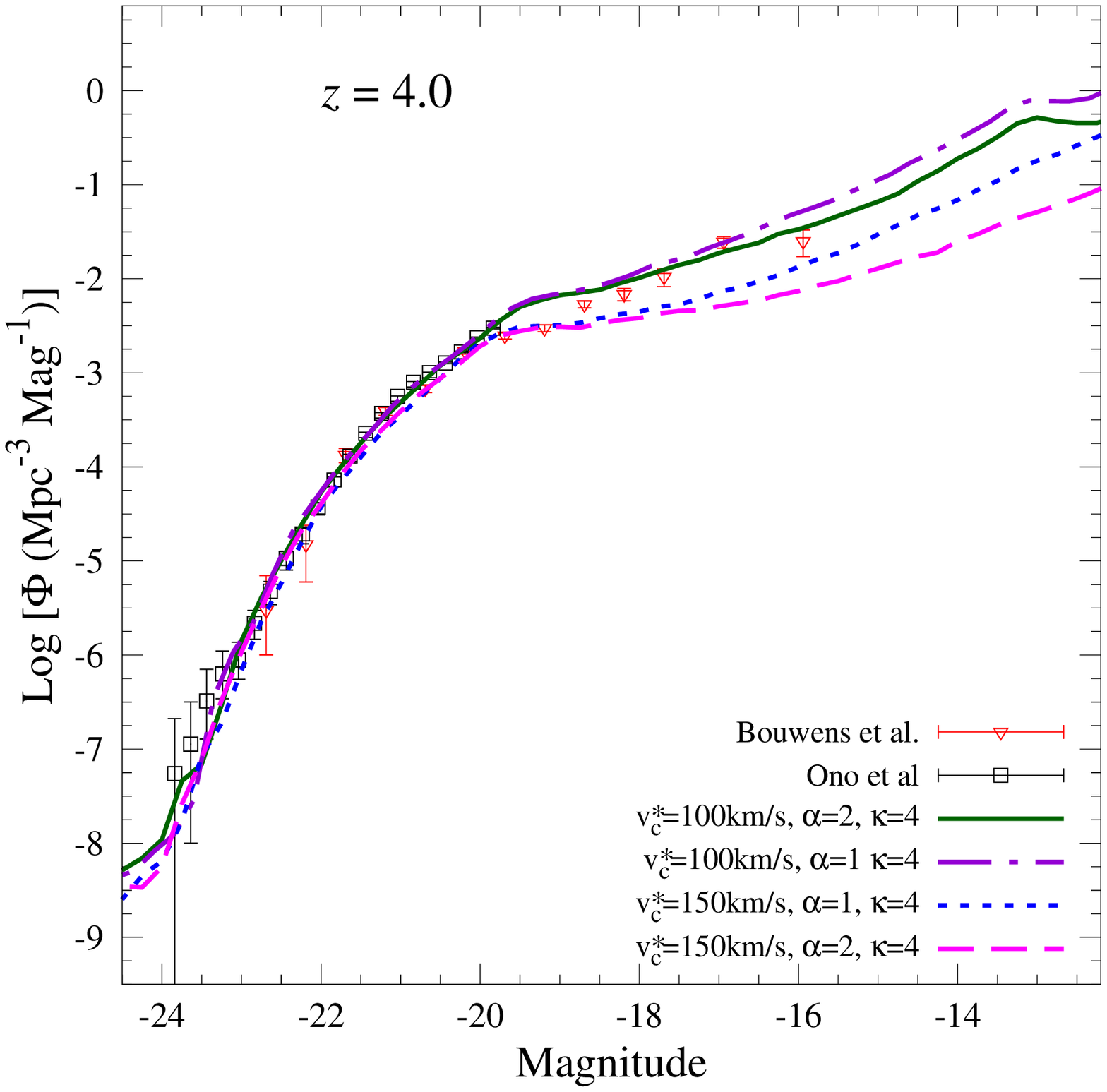,width=5.5cm,angle=-90.}%
\epsfig{figure=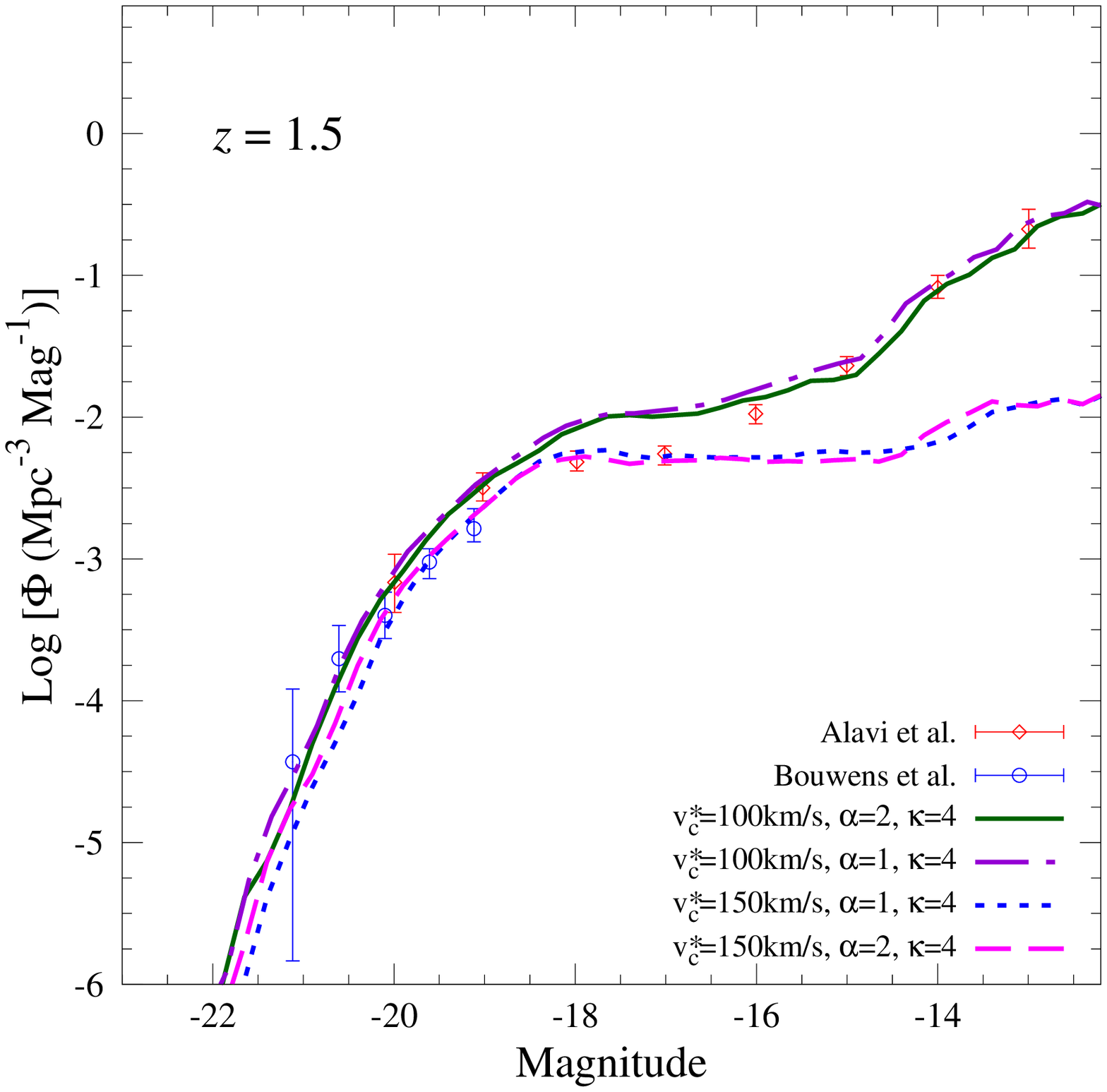,width=5.5cm,angle=-90.}%
}
\caption[]{Comparison of luminosity functions predicted by models with different $v_c^*$ at $z=8$ (left panel),
4 (middle panel) and 1.5 (right panel). Model parameters are described in legends.
}
\label{fig_model_comp}
\end{figure*}

In previous sub-sections we have demonstrated the importance of SNe feedback in understanding
the shape, extent and the redshift evolution of the UV luminosity function of galaxies.
Here we discuss the effect of the parameter $v_c^*$ that governs the
level of SNe feedback on the luminosity function.
In Fig.~\ref{fig_model_comp} we have shown predicted luminosity functions with models having $v_c^*=150$~km/s
(dotted black lines for $\alpha=1$ and magenta small dashed lines for $\alpha=2$)
along with two other models that were discussed in previous sections
(with $v_c^*=100$~km/s). We show luminosity functions only for three representative redshifts,
$z=8$, 4 and 1.5. It is clear from the figure that even though models with $v_c^*=150$~km/s are degenerate
at the level of observational uncertainty with our fiducial model at $z=8$, the difference becomes more
and more as we come to lower and lower redshifts. At redshift $z=1.5$ the observations clearly rule out
possibility of having a higher $v_c^*$. Note that \citet{2010MNRAS.402.2778S} obtained $v_c^* \sim 100$~km/s
for outflows driven by both hot gas and cosmic rays.

\section{Discussions and Conclusions}
\label{sec_dc}

Over the past decade
we have been developing semi-analytical models of galaxy formation
which start from the abundance of dark matter halos,
include a simple prescription for star formation and different feedback processes
to predict various observed properties of high-$z$
galaxies. 
In particular our models predict some specific observational signatures: (i) turn over in the
low luminosity end of the LF due to cooling criteria adopted, (ii) faint end slope of the LF that depends
on the SNe feedback and radiative feedback from the meta-galactic UV background,
and (iii) bright end slope of the LF  dominated by the AGN feedback.
All our models are tuned to satisfy the reionization constraints obtained
from CMB and Lyman-$\alpha$ forest observations.

Latest advancements in observations now provide UV luminosity
functions of high-$z$ galaxies over a large luminosity range and allow
us to check the above mentioned predictions. 
In this paper we have compared the latest measurements of the UV luminosity
functions at different epochs (i.e $1.5\le z\le8$) with the predictions of
our semi-analytical galaxy formation models and draw the following conclusions.

\begin{itemize}
\item{} In order to reproduce the observed redshift evolution of the LF, our models require the parameter,
  $f_*/\eta$, to decrease with decreasing redshifts. By using the observed redshift evolution
  of UV extinction parameter, A$_{\rm FUV}$,  to constrain $\eta(z)$ we conclude that the conversion
  efficiency of gas into stars is decreasing with decreasing redshift.

\item{} We show that SNe feedback is important even at $z>6$ to reproduce the
  observed low end of the LF. Radiative feedback alone will not be sufficient to
  reproduce the observations. We show precise measurements of LF in future will
  enable us to distinguish between different modes of SNe feedback (i.e momentum
  or energy driven wind feedback), duration of the star formation ($\kappa$) and circular
  velocity scale ($v_c^*$) where we normalise the mass outflow rate. 
  Our best fit models that consider star formation only in the ``atomic cooled halos''
  predict a turn over in the low end of the LF of high-$z$ galaxies (i.e $z>6$)
  at M$_{\rm UV}\sim-13$. 
  This is much below the limit reached by present day observations.
  At these redshifts, our models also produce the LF in the high luminosity end very well even when
  AGN feedback is not included.

\item{} Our models with the above noted redshift evolution in $f_*/\eta$
  reproduce the observed LF at the intermediate redshifts ($3\le z \le 5$)
  very well.
  We also present our model predictions at the low luminosity end that is not yet probed by the
  present observations. In particular, the presence of a passive mode of star formation in
  our models predicts excess of galaxies at low luminosity end. Detecting such an excess will
  allow us to place constraints on such a slowly evolving populations at high-$z$. In this
  redshift range in order to reproduce the LF at high luminosity end we do need AGN feedback.

\item{} The observed luminosity functions at low-$z$ (i.e $1.5\le z\le 2.5$) show an upward turn
  at $M_{\rm UV}>-15$. This part of the LF in our models is produced by galaxies exhibiting slow
  passive mode of star formation. It will be  possible to confirm this by a detailed
  SED fitting of the multiband photometric data of these galaxies. In our models, these
  galaxies typically have halo masses in the range $10^9\le M(M_\odot) \le 10^{11}$.
  Further we find the characteristic velocity ($v_c^*$)  at which the mass loading factor
  $\eta_w = 1$ can be constrained using the LF.
  
\end{itemize}

In summary, we show that the observed luminosity functions in the entire
redshift range consider here i.e. $1.5 \le z \le 8$ show signature
of galaxies going through the prolonged slow mode of star formation in addition
to star bursting galaxies.
We also demonstrate that deep observations
of faint galaxies at high redshift will enable us to probe the nature of feedback
such as momentum/energy driven flow, mass range over which the radiative feedback
is effective, normalization of mass outflow rates and nature of galaxies going through
passive mode of star formation. At present one is able to detect faint galaxies
thanks to the gravitational lensing by the foreground clusters. However, the
luminosity function derived using this technique suffers from systematic uncertainties
associated with the lens modelling of the system. This will be removed once we get
direct observations of faint galaxies using future large observing facilities like
JWST, TMT and ELT.

\section*{Acknowledgements}
SS thanks IUCAA for its support through associateship programme.

\bibliography{Faint_LF}
\bsp	
\label{lastpage}

\end{document}